\documentclass[aps,prb,notitlepage,twocolumn]{revtex4-1}
\usepackage{amsmath,amssymb,amsfonts,graphicx,subfigure,xcolor}

\def\bea{\begin{eqnarray}}
\def\eea{\end{eqnarray}}
\def\nn{\nonumber}
\def\ba{\begin{array}}
\def\ea{\end{array}}
\def\nn{\nonumber}

\def\O{\mathcal{O}}

\def\I{\mathcal{I}}

\def\ii{\mathfrak{i}}
\def\tb{\textbf}
\def\arccoth{\text{arccoth}}
\def\Im{\text{Im}}

\begin{document}

\title{Hydrodynamic sound and plasmons in three dimensions}

\author{Shao-Kai Jian}
\author{Sankar Das Sarma}
\affiliation{Condensed Matter Theory Center and Joint Quantum Institute, Department of Physics, University of Maryland, College Park, Maryland 20742, USA}

\begin{abstract}
	In a recent paper by Lucas and Das Sarma [Physical Review B {\bf97}, 115449 (2018)], a solvable model of collective modes in 2D metals was considered in the hydrodynamic regime. In the current work, we generalize the hydrodynamic theory to 3D metals for which the calculation of sound modes in a strongly-coupled quantum Coulomb plasma can be made explicit. The specific theoretical question of interest is how the usual linearly dispersing hydrodynamic sound mode relates to the well-known gapped 3D plasmon collective mode in the presence of long-range Coulomb interaction. We show analytically that both the zero sound in the collisionless regime and the first sound in the hydrodynamic region become massive in 3D, acquiring a finite gap because of the long-range Coulomb interaction, while their damping rates become quadratic in momentum. We also discuss other types of long-range potential, where the dispersion of sound modes is modified accordingly. The general result is that the leading order hydrodynamic sound mode is always given by the leading-order plasmon frequency in the presence of long-range Coulomb interaction, but the next-to-leading-order dispersion corrections differ in hydrodynamic and  colllisionless regimes.
\end{abstract}


\maketitle

\parskip=10pt

\section{Introduction}

An interacting electron system could be in the strongly interacting collision-dominated hydrodynamic regime where frequent inter-particle collisions lead to local thermodynamic equilibrium so that the system should be thought of as manifesting a collective hydrodynamic behavior rather than the usual collisionless behavior of weak interactions where the standard independent quasiparticle picture applies. The analogy is to real liquids (e.g. water) whose macroscopic long wavelength flow properties hardly depend on the microscopic details of the molecular interactions which only serve to determine the macroscopic hydrodynamic parameters such as viscosity. Although the possibility of electron hydrodynamics was suggested a long time ago~\cite{Gurzhi:1968hydrodynamic}, there has been a great deal of recent interest in the subject arising from the possibility of the experimental observation of electron hydrodynamics in solid state materials~\cite{Lucas:2018hydrodynamics}. The current work focuses on a specific theoretical question regarding electron hydrodynamics.

The question we address is the interplay of the electronic plasmon mode with the hydrodynamic sound mode: How does the plasmon affect the hydrodynamic sound mode and does the electron system in the hydrodynamic regime undergo collective plasmon oscillations at all? 

The theoretical question was recently addressed for 2D metals in a recent work~\cite{Lucas:2018electronic}, and we generalize the theory to 3D metals. The 3D generalization is nontrivial and is of interest because of the true long-range nature of electronic Coulomb interaction in 3D metals, leading to a collective plasmon mode with a finite energy even in the long wavelength limit~\cite{Bohm:1953collective}. This gapped massive nature of 3D plasmons arises from the 3D Coulomb coupling going as $1/q^2$, where $q$ is the wavenumber (or momentum). By contrast, 2D Coulomb coupling goes as $1/q$, leading to the 2D plasmon going at long wavelength as $q^{1/2}$. Since the hydrodynamic sound mode (the so-called 'first sound' in the helium literature), which is the electron analog of the ordinary acoustic sound, goes linear in $q$ at long wavelength by definition, the interplay of a gapped 3D plasmon going as $\O(q^0)$ with hydrodynamic sound going as $\O(q)$ is more intriguing than the interplay between the hydrodynamic sound and the 2D plasmon, both of which vanish at long wavelength albeit with different momentum scaling. If the hydrodynamic sound mode in 3D metals develops a long wavelength gap by virtue of Coulomb interaction, this becomes reminiscent of the Higgs mechanism with a linearly dispersing Goldstone mode acquiring a mass by virtue of long-range coupling although the electron hydrodynamics problem does not involve any symmetry breaking (or an underlying Higgs field) in order to produce the gapped sound mode. In fact, this is precisely what happens in a metallic superconductor where the expected linearly dispersing Goldstone mode associated with the spontaneous breaking of the gauge symmetry acquires a long wavelength gap becoming effectively the plasmon mode in the presence of long-range Coulomb coupling instead of the usual zero sound acoustic mode as in a neutral superfluid where there is no long-range Coulomb coupling~\cite{Anderson:1958random, Anderson:1963plasmons, Prange:1963dielectric, Fertig:1990collective, Fertig:1991collective}.

We show in this work that indeed 3D Coulomb coupling leads to a mass or a gap in the hydrodynamic sound mode in 3D metals, and the sound mode becomes the effective long wavelength plasmon mode in the hydrodynamic regime of 3D metals.  This is, however, true only in the leading order in momentum, and the next-to-leading-order dispersion corrections in wavenumber are different for 3D plasmons in the collisionless regime and the first sound in the hydrodynamic regime. We also study the damping or decay of the plasmon mode (which is akin to the zero sound mode in the helium literature) in the collisionless regime and the hydrodynamic first sound mode in the collision-dominated regime. Our terminology in the paper uses 'collisionless' to imply the non-hydrodynamic regime of weak inter-particle collisions (where the standard 'plasmon' or zero sound mode exists). By contrast, the collision-dominated regime is the hydrodynamic regime of rapid inter-particle collisions where the first sound mode exists. We study both regimes including effects of 3D long-range Coulomb interaction.

The hydrodynamic description applies when the momentum conserving inelastic electron-electron interaction is much stronger than any other momentum relaxing elastic scattering mechanisms which might be present in the system. In metals, such momentum non-conserving scattering processes arise from electron-impurity and electron-phonon scattering, which typically dominate at low and high temperatures respectively, making the hydrodynamic regime difficult to realize experimentally in the laboratory although, in principle, a very clean metal should manifest electron hydrodynamics at very low temperatures where the phonon scattering rate (in terms of resistivity) is suppressed as $T^5$ and the quasiparticle scattering rate goes as $T^2$, where $T$ is the temperature~\cite{Ashcroft:1976solid}. Eventually, at low enough temperatures hydrodynamics is cut off in metals by any residual impurity scattering. In a metal with negligible impurity and phonon scattering, electron-electron interactions should produce hydrodynamic behavior at long wavelength and low frequency. 

\section{Summary of the main results} \label{sec:summary}

We extend the theory of~\cite{Lucas:2018electronic} to three dimensions. Namely, we construct a solvable model that admits the exact calculation of sound modes in both the collisionless regime and hydrodynamic regime. The sound mode in the collisionless regime is the zero sound mode or the plasmon, and the sound mode in the hydrodynamic regime is the usual sound mode (i.e., the first sound). In this section, we represent the main results from the solvable model, including the effect of long-range Coulomb interaction on the sound modes. We leave the detailed derivation of the results to the following sections, but the exactly solvable model already demonstrates the physics clearly.

In this exactly solvable model consisting of spinless fermions, we assume spherical symmetry and consider the nonvanishing Landau parameter only in the s-wave channel. We denote the dimensionless Landau parameter $F_0$ and we consider repulsive interactions so that $F_0>0$. The sound mode is highly damped when the interaction is attractive, and may even lead to instabilities~\cite{Nozieres:2018theory}. In any case, our interest is 3D metals in which the direct electron-electron interaction is repulsive. These considerations make the calculations analytically tractable, while maintaining the essential physics. 

We first present the result in the clean limit where the sound mode is not damped by impurity scattering, $\gamma_\text{imp} = 0$.  $\gamma_\text{imp}$ is the scattering rate between quasiparticles and impurities. The zero sound in the collisionless regime $\omega \gg \gamma$, where $\gamma$ denotes the scattering rate between quasiparticles, is given by
\bea \label{eq:zeroFull}
&& \omega = \pm v_0 q - \ii \gamma \frac{(F_0+1)^2 - 2(F_0-1) (\frac{v_0}{v_F})^2 - 3 (\frac{v_0}{v_F})^4 }{F_0 [ F_0 +1 - (\frac{v_0}{v_F})^2]}, \\  && \frac{v_0}{v_F} \arccoth \frac{v_0}{v_F} = 1 + \frac1{F_0}.
\eea
where the second equation determines the zero sound velocity. This equation is valid for all $F_0>0$. For the weakly interacting Fermi liquid, $F_0 \ll 1$, and the zero sound velocity is nonperturbative in interaction strength, namely, $v_0 = v_F (1 + e^{-2/F_0})$. It approaches the Fermi velocity as one should anticipate for the free electron gas. For $F_0 > 1$, the dispersion can be simplified as
\bea \label{eq:zero}
		\omega = \pm \sqrt{\frac{F_0}3 + \frac35} v_F q - \ii \gamma \frac{2(5F_0 + 21)}{5F_0(5F_0 + 3)},
\eea
where $v_F$ is the Fermi velocity. Decreasing the frequency, the system enters the collision-dominated hydrodynamic regime where $ \omega \ll \gamma$. Then the zero sound crosses over smoothly to the first sound,
\bea \label{eq:first}
\omega = \pm \sqrt{ \frac{F_0}3 + \frac13} v_F q - \ii \frac{2 v_F^2}{15 \gamma} q^2.
\eea

\begin{figure}
	\centering
	\includegraphics[width=6cm]{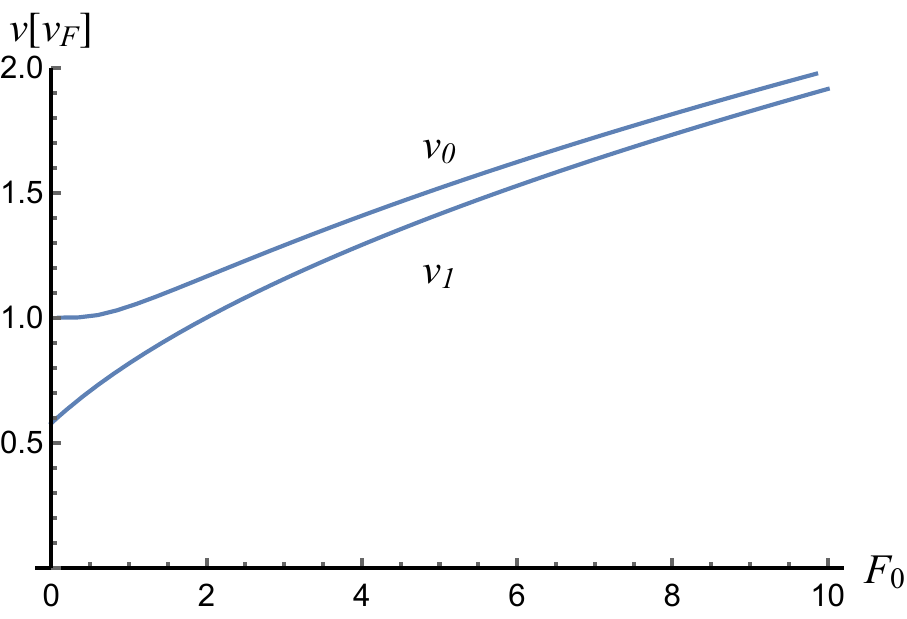} 
	\caption{\label{fig:velocity}The zero sound and first sound velocities as a function of the Landau parameter, where the Fermi velocity is set to 1 as the unit. The zero sound velocity is larger than the first sound velocity for all $F_0>0$.}
\end{figure}
The linear term in momentum reveals the sound velocity. In Fig.~\ref{fig:velocity}, one can see that the collisionless zero sound velocity is always larger than the hydrodynamic first sound velocity, $v_0>v_1$. For the asymptotically large Landau parameter, the following inequality holds true, too, i.e.,
\bea
\frac{v_0}{v_1} = \sqrt{\frac{F_0 + 9/5}{F_0+1}} > 1 . 
\eea
The imaginary part leads to the damping of sound modes. In an interacting Fermi liquid, the scattering rate at low temperatures is given by $\gamma \propto T^2$. A crucial difference between the zero sound and the first sound is that the damping rate is proportional to the interaction scattering rate, $\Im(\omega) \propto\gamma$, for zero sound while it is inversely proportional to the interaction scattering rate, $\Im(\omega) \propto \gamma^{-1}$, for the first sound. These results are well known and also experimentally observed in normal He-3~\cite{Abel:1966propagation}. Reproducing these results presents a consistency check of our solvable model.

When the impurity scattering is strong $\gamma_\text{imp} > v_F q$, both sound modes will be damped by impurity scattering at small momentum. The propagating wave transitions to a quadratic diffusion mode
\bea
	\omega = - \ii \frac{v_1^2}{\gamma_\text{imp}}  q^2, \quad v_1 = v_F \sqrt{\frac{1+F_0}3}
\eea
and a fully gapped one $ \omega = -\ii \gamma_\text{imp}$. When the impurity scattering is weak $\gamma_\text{imp} < v_F q$, its effect is an additional damping of the zero sound mode (in addition to the damping induced by quasiparticle interactions $\gamma$); that is, the correction is $ \delta\Im(\omega) = \frac12 \gamma_\text{imp}$, also see~(\ref{eq:weak_imp}). 

Now we discuss the effect of Coulomb interaction on the sound mode. It is well known that the plasmon is fully gapped in 3D metals because of Coulomb interaction~\cite{Bohm:1953collective}. Since both the plasmon and the sound wave are density fluctuating collective modes in many-body systems, it is naturally expected that the sound mode should also develop a finite gap due to the Coulomb interaction. Since Coulomb interaction acts at the s-wave channel, one can make the following replacement,
\bea
	F_0 \rightarrow F_0 + \frac{8\pi \alpha}{\lambda_F^2 q^2},
\eea
where $\alpha = \frac{e^2}{4\pi v_F}$ is the effective finite structure constant for the 3D metal ($v_F$ is the Fermi velocity), and $\lambda_F = \frac{2\pi}{k_F}$ is the Fermi wavelength.
The Coulomb interaction [the second term in (6)] should be understood as the effective one felt by quasiparticles, where the effective fine structure constant captures all possible renormalization effects in going from bare electrons to quasiparticles. This is a generalization from the neutral Fermi liquid theory to the Landau-Silin theory of a charged Fermi liquid~\cite{Zala2001:interaction}.
Since we are interested in the long wavelength limit, where the Coulomb interaction dominates over any short-range interaction, we focus on~(\ref{eq:zero}) in the appropriate limit. After making this replacement, the zero sound becomes
\bea \label{eq:zeroPlas}
	\omega = \pm \left( \sqrt{\frac{8\pi \alpha}{3}} \frac{v_F}{\lambda_F} + \frac{9/5+F_0}{4\sqrt{6\pi \alpha}} v_F \lambda_F q^2 \right) - \ii \gamma \frac{\lambda_F^2 q^2}{20 \pi \alpha},
\eea
and the first sound becomes
\bea \label{eq:firstPlas}
	\omega = \pm \left( \sqrt{\frac{8\pi \alpha}{3}} \frac{v_F}{\lambda_F} + \frac{1+F_0}{4\sqrt{6\pi \alpha}} v_F \lambda_F q^2 \right) - \ii \frac{2 v_F^2}{15 \gamma} q^2.
\eea

It is instructive to compare the results with the plasmon mode. The 3D plasmon dispersion is well known and is determined by the following equation in RPA calculations,
\bea \label{eq:plasmon3D}
	\omega^2 = \omega_0^2 + \frac35 \frac{k_F^2}{m^2} \frac{\omega_0^2}{\omega^2} q^2, \quad \omega_0^2 = \frac{e^2 N_e}{m},
\eea
where $N_e = \frac{4\pi}3 \frac{k_F^3}{(2\pi)^3} = \frac{k_F^3}{6\pi^2}$ is the electron density in 3D,  $m = \frac{k_F}{v_F}$ is the effective mass of the quasiparticle, and $\omega_0$ is the well-known plasmon frequency. (Note a conventional difference between the plasmon frequency in our~(\ref{eq:plasmon3D}) and Eq.~(14) in Ref.~[4]
, which comes from the Coulomb potential we define as~(\ref{eq:energy}) having an extra factor of $1/4\pi$ in real space as we use the rationalized unit instead of the Gauss unit.) If we expand the plasmon mode at long wavelength, its dispersion reads
\bea \label{eq:3Dplasmon}
	\omega = \omega_0 + \frac35 \frac{k_F^2}{m^2 \omega_0} \frac{q^2}2 =  \sqrt{\frac{8\pi \alpha}{3}} \frac{v_F}{\lambda_F} + \frac{9/5}{4\sqrt{6\pi \alpha}} v_F \lambda_F q^2 , \nn \\
\eea
where in the second step, we change the parameters to better compare with sound mode results. Comparing~(\ref{eq:3Dplasmon}) to the zero sound mode with Coulomb interaction~(\ref{eq:zeroPlas}), we find a correction to the quadratic dispersion from the Landau parameter $F_0$ and a quadratic damping due to the collisions. Comparing~(\ref{eq:3Dplasmon}) to the first sound mode with Coulomb interaction~(\ref{eq:firstPlas}), we conclude that in the hydrodynamic regime, the plasmon dispersion gets corrected by a factor of $5/9$ at the next-to-leading $q^2$ order because of Coulomb coupling. Again there is a damping effect in~(\ref{eq:firstPlas}) inherited from the first sound mode.

More generally, we can consider a formal long-range interaction given by a power law defined by $q^{-2\eta}$, where $\eta=1$ for 3D Coulomb coupling, i.e.,
\bea
	F_0 \rightarrow F_0 + \frac{8\pi \alpha}{(\lambda_F q)^{2\eta}},
\eea
where $\alpha$ is the effective interaction strength, and the parameter $\eta$ defines the form of the interaction. In real space, this translates to a potential of the form $r^{2\eta - 3}$. For $\eta >1$ ($\eta < 1$) it is stronger (weaker) than the Coulomb potential. This type of interaction leads to the zero sound mode
\bea
\omega = \pm \sqrt{\frac{8\pi \alpha}{3}} \frac{v_F}{\lambda_F^\eta} q^{1-\eta} - \ii \gamma \frac{(\lambda_F q)^{2\eta}}{20 \pi \alpha},
\eea
and the first sound mode
\bea
\omega = \pm \sqrt{\frac{8\pi \alpha}{3}} \frac{v_F}{\lambda_F^\eta} q^{1-\eta} - \ii \frac{2 v_F^2}{15 \gamma} q^2.
\eea
In this case, the sound mode has an interesting dispersion resulting from the long-range interaction. Also note that the damping rate for the "hydrodynamic plasmon" does not change due to the specific form of the long-range interaction, remaining independent of the range parameter $\eta$.

\section{Review of the Boltzmann equation in a 3D metal}

The collective mode is one of the fundamental excitations of a many-body system. It emerges from coherent interactions between quasiparticles, and is fundamentally different from single particle electron-hole type excitations. It is convenient to describe the collective mode by the distribution function $n(\tb k, \tb r, t)$ of the quasiparticle at given momentum $\tb k$ and position $\tb r$. As we are interested in the effect of Coulomb interaction on the sound mode, we will restrict ourselves to the spinless electron. Spin can be added straightforwardly at the cost of making the notations cumbersome---we emphasize that the modes we are discussing are charge density collective excitations which are independent of electron spin. The Boltzmann equation governing the dynamics of the distribution function is~\cite{Nozieres:2018theory}
\bea
	\frac{d n(\textbf k, \textbf r, t)}{dt} = \left( - \frac{d \textbf r}{dt} \cdot \frac{\partial}{\partial \textbf r} - \frac{d \textbf k}{dt} \cdot \frac{\partial}{\partial \textbf k} \right) n(\textbf k, \tb r, t) + \I[n], \nn \\
\eea
where the first term on the right hand side is the drift term while the second term is the collision integral. The semiclassical equation of motion of a quasiparticle is well known
\bea
	\frac{d \tb k}{d t} &=& - \frac{\partial \epsilon(\tb k, \tb r, t)}{\partial \tb r}, \\
	\frac{d \tb r}{d t} &=&  \frac{\partial \epsilon(\tb k, \tb r, t)}{\partial \tb k},
\eea
where $\epsilon(\tb k, \tb r, t)$ is the energy of the quasiparticle. Since we are interested in a conventional metal, it is sufficient to consider the semiclassical description without external electromagnetic field or Berry curvature. The quasiparticle energy should be determined self-consistently from the Boltzmann equation. We are going to solve the collective mode of a small variation from the Fermi-Dirac distribution function, namely,
\bea
	n(\tb k, \tb r, t) = n_F(\tb k) + \delta n(\tb k, \tb r, t), \\
	n_F(\tb k) \equiv n_F[\epsilon_0(\tb k)] = \frac1{e^{\beta(\epsilon_0(\tb k) -\mu)}+1},
\eea
where $n_F(\tb k)$ is the Fermi-Dirac distribution at equilibrium, $\epsilon_0(\tb k)$ is the bare energy of free electrons, $\beta \equiv 1/T$ is the inverse temperature, and $\mu$ denotes the chemical potential.

The total energy of the system with a small variation from equilibrium is
\bea 
	\epsilon_{\text{tot}}(t) &=& \int d^3 \tb r \int_{\tb k} \epsilon_0(\tb k) \delta n(\tb k, \tb r, t) \nn\\
	&& + \frac12 \int d^3 \tb r \int_{\tb k, \tb k'} \delta n(\tb k, \tb r, t) f(\tb k, \tb k') \delta n(\tb k', \tb r, t) \nn\\
\label{eq:energy}	&& + \frac12 \int d^3 \tb r d^3 \tb r' \int_{\tb k, \tb k'} \delta n(\tb k, \tb r, t) \frac{e^2}{4\pi}\frac{1}{|\tb r - \tb r'|} \delta n(\tb k', \tb r', t), \nn\\
\eea
where $\int_{\tb k} \equiv \int \frac{d^3 \tb k}{(2\pi)^3}$ and $f(\tb k, \tb k')$ is the Landau parameter characterizing the short-range quasiparticle interactions, and the second line represents the long-range Coulomb interaction. 
The presence of both short-range and long-range interactions is an important generalization to a charged Fermi liquid from the neutral Fermi liquid theory. 
This generalization is also often referred to as Landau-Silin Fermi liquid theory.
Note that throughout the paper, we use the rationalized unit, where the factor of $\frac1{4\pi}$ appears in the real space Coulomb potential.
Thus, by varying with respect to $\delta n(\tb k, \tb r, t)$, we can get the quasiparticle energy
\bea
	\epsilon(\tb k, \tb r, t) =  \epsilon_0(\tb k) + \int_{\tb k'} f(\tb k, \tb k') \delta n(\tb k', \tb r, t) \nn \\
	+ \int_{\tb k'} \int d^3 \tb r' \frac{e^2}{4\pi} \frac{1}{|\tb r - \tb r'|} \delta n(\tb k', \tb r', t).
\eea

To get a wave-like collective mode, we assume in the usual manner that the variation takes the plane wave form
\bea
	\delta n(\tb k, \tb r, t) = \delta n(\tb k) e^{\ii \tb q \cdot \tb r - \ii \omega t}.
\eea
Actually, once we get the eigenmode from the plane wave expansion, we can construct any arbitrary mode using linear superposition and completeness. The energy for such a plane wave excitation is
\bea
	\epsilon(\tb k, \tb r, t) &=& \epsilon_0(\tb k) + \int_{\tb k'} f(\tb k, \tb k') \delta n(\tb k') e^{\ii \tb q \cdot \tb r - \ii \omega t} \nn\\ 
	&& + \int_{\tb k'} \int d^3 \tb r' \frac{e^2}{4\pi}\frac{1}{|\tb r - \tb r'|} \delta n(\tb k') e^{\ii \tb q \cdot \tb r' - \ii \omega t} \\
	&=& \epsilon_0(\tb k) + \int_{\tb k'} \left( f(\tb k, \tb k') +\frac{e^2}{\tb q^2} \right) \delta n(\tb k') e^{\ii \tb q \cdot \tb r - \ii \omega t} , \nn\\
\eea
where in the second line, we have used the Fourier transform of the Coulomb potential in 3D. A simple derivation is given in Appendix~\ref{append:Coulomb}. 

Now with the quasiparticle energy, the semiclassical equation of motion becomes
\bea
	\frac{d \tb k}{dt} &=& - \ii \tb q \int_{\tb k'} \left( f(\tb k, \tb k') + \frac{e^2}{\tb q^2}  \right) \delta n(\tb k') e^{\ii \tb q \cdot \tb r - \ii \omega t}, \\  \frac{d \tb r}{dt} &=& \frac{\partial \epsilon_0(\tb k)}{\partial \tb k} \equiv \tb v(\tb k).
\eea
Putting this equation of motion into the Boltzmann equation, we arrive at
\bea
	\omega \delta n(\tb k) = \tb q \cdot \tb v(\tb k) \Big( \delta n(\tb k) - \big[\partial_\epsilon\big|_{\epsilon = \epsilon_0(\tb k)} n_F(\epsilon) \big] \nn \\
	\times \int_{\tb k'} \big(f(\tb k, \tb k') + \frac{e^2}{\tb q^2} \big) \delta n(\tb k') \Big) + \ii \I[n],
\eea
where $\partial_\epsilon\big|_{\epsilon = \epsilon_0(\tb k)}$ means taking the derivative with respect to $\epsilon$ and then setting $\epsilon= \epsilon_0(\tb k)$. Since at low temperatures, the small variation $\delta n(\tb k)$ concentrates near the Fermi surface according to the factor 
\bea
	\lim_{\beta \rightarrow \infty} -\partial_\epsilon\big|_{\epsilon = \epsilon_0(\tb k)} n_F(\epsilon) =\delta\big(\epsilon_0(\tb k) - \mu\big),
\eea
which tends to a delta function localized at the Fermi surface, it is easy to see that the solution to the equation has the form $\delta n(\tb k) = - [\partial_\epsilon|_{\epsilon = \epsilon_0(\tb k)} n_F(\epsilon)] \delta n(\sigma) = \delta(\epsilon_0(\tb k) - \mu) \delta n(\sigma)$. Here, $\sigma = (\theta, \phi) $ is determined by the vector $\tb k$ at the Fermi surface. Hence, it is convenient to change the variable from $\tb k$ to $\big( \epsilon_0(\tb k), \sigma \big) $. With the help of the identity,
\bea
	\int d^3 \tb k = \int_0^\infty d \epsilon \int_{\epsilon_0(\tb k) = \epsilon} \frac{d\sigma}{|\tb v(\tb k)|}, 
\eea
where $d \sigma$ denotes the measure over the Fermi surface, and $d \epsilon$ denotes the measure perpendicular to the Fermi surface, we arrive at
\bea \label{eq:Boltzmann}
	&& \omega \delta n(\sigma) = \tb q \cdot \tb v_F(\sigma) \Big( \delta n(\sigma) \nn \\ 
	&& + \frac1{(2\pi)^3} \int \frac{d\sigma'}{|\tb v_F(\sigma')|} \big(f(\sigma, \sigma')  + \frac{e^2}{\tb q^2} \big) \delta n(\sigma') \Big) + \ii \I[n],
\eea
where due to the delta function, the integration is restricted to the Fermi surface, and we use $\tb v_F(\sigma)$ to denote $\tb v(\tb k)$ when $\tb k$ is located at the Fermi surface which is the Fermi velocity, and we also use $f(\sigma, \sigma')$ to denote $f(\tb k, \tb k')$ when both $\tb k$ and $\tb k'$ are on the Fermi surface. The Coulomb interaction is independent of the angle variable and $\tb q$ is not a dynamical quantity, so one can regard the Coulomb interaction as a modification of the Landau parameter in the s-wave channel.

The Boltzmann equation~(\ref{eq:Boltzmann}) is the central equation that governs the collective modes in a Fermi liquid, including sound modes. It can describe the situation with either short-range interaction or long-range interaction, treating the zero sound, first sound, and plasmon equivalently within one formalism. In the next section, we construct a simple model where the Boltzmann equation~(\ref{eq:Boltzmann}) can be solved exactly.

\section{A solvable model}

To proceed, let us assume the Fermi liquid in question has spherical symmetry. This is the situation in simple 3D metals. As a result the Fermi velocity is independent of angle but $|\tb v_F(\sigma)|= v_F$ and we can choose $\tb q = (0,0, q)$ pointing along the $k_z$ direction, and use the spherical coordinate $\Omega=(\theta, \phi)$. Then (\ref{eq:Boltzmann}) becomes
\bea \label{eq:Boltzmann_sph}
	&& \omega \delta n(\Omega) = q v_F \cos\theta \Big( \delta n(\Omega) \nn\\
	&& + \int \frac{d\Omega'}{4\pi}\left( F(\Omega, \Omega') + \frac{8\pi \alpha}{\lambda_F^2 \tb q^2} \right) \delta n(\Omega') \Big) + \ii \I[n],
\eea
where we have used $\int d \sigma = k_F^2 \int d\Omega = k_F^2 \int \sin \theta d\theta d \phi$, $k_F $ is the Fermi momentum $\epsilon_0(k_F) =\mu $ and $\lambda_F$ is the corresponding Fermi wavelength $\lambda_F = \frac{2\pi}{k_F}$. $\alpha = \frac{e^2}{4\pi v_F}$ is the effective fine structure constant in the Fermi liquid defining the interaction coupling strength. 
$F(\Omega, \Omega') \equiv \frac{k_F^2 }{2\pi^2 v_F} f(\Omega, \Omega')$ is the dimensionless Landau parameter arising from changing the variables from momenta to angles at the Fermi surface.
As we mentioned in the previous section, the Coulomb interaction is not a dynamical quantity in the Boltzmann equation. We can absorb the Coulomb interaction into the Landau parameter in the s-wave channel, and restore it back at the end of the calculation. 

In the following, we assume that the only nonvanishing component of the Landau parameter is in the s-wave channel, namely, the Landau parameter is a constant $F(\Omega, \Omega') = F_0 $. We absorb the Coulomb interaction into the Landau parameter. This should be sufficient for our purpose to investigate the effect of the Coulomb interaction on the sound mode in a solvable mode. Without the collision integral, the Boltzmann equation reduces to the following eigen equation,
\bea \label{eq:Boltzmann_zero} 
	(x_0 - \cos \theta) \delta n(\Omega) = F_0 \int \frac{d\Omega'}{4\pi}  \delta n(\Omega') , \quad x_0 = \frac{\omega}{q v_F},
\eea
which can solved~\cite{Nozieres:2018theory} by 
\bea \label{eq:n0}
\delta n(\Omega) = \frac{\cos \theta}{x_0 - \cos \theta}, \quad x_0 \arccoth x_0  = 1+  \frac1{F_0}.
\eea
Since we ignore the collision integral, this solution represents, by definition, the zero sound solution in the collisionless regime. The second equation determines the velocity of the zero sound. When the $F_0 >1$, we get the approximate zero sound velocity given by $v_0 = \sqrt{\frac{F_0}3 + \frac35} $.

Since (\ref{eq:Boltzmann_zero}) has spherical symmetry, we can study the eigen equation using the spherical harmonics. Any solution $\delta n(\Omega)$ can be expanded in the basis of spherical harmonics,
\bea
	\delta n (\Omega) = \sum_{l=0}^\infty \sum_{m=-l}^l \delta n_{lm} Y_l^m(\theta, \phi),
\eea
where $Y_l^m(\theta, \phi)$ is the spherical harmonic and $\delta n_{lm}$ is the corresponding expansion coefficient. The sound modes are solutions with zero magnetic quantum number $m=0$. As the eigen equation conserves the magnetic quantum number, we consider the $m=0$ sector where the equation can be cast into
\bea
	x_0 \delta n_{l,0} &=& \frac{l}{\sqrt{4l^2-1}} \delta n_{l-1,0} + \frac{l+1}{\sqrt{4(l+1)^2-1}} \delta n_{l+1,0} \nn\\ && + \frac{F_0}{\sqrt3} \delta n_{0,0} \delta_{l,1}.
\eea
The detailed derivation of this equation is given in Appendix~\ref{append:math}. For $l \ge 2$, it resembles a recurrence relation of a series. Indeed, it is not hard to check that for $l\ge 2$, the following series of hypergeometric functions satisfies the recurrence relation,
\bea \label{eq:series}
	a_{l,m} (x) &=& 2\pi \delta_{m,0} \sqrt{\frac{2l+1}{4}}  \frac{1}{(2x)^l} \frac{\Gamma(l+1)}{\Gamma(l+ \frac32)} \nn\\
	&&\times \ _2F_1\left( \frac{l+1}2 , \frac{l+2}2; l + \frac32; \frac1{x^2}\right), ~ l \ge 1.
\eea
This series is consistent with the eigenfunction (\ref{eq:n0}). We present the details of obtaining the series in Appendix~\ref{append:math}. Thus, we have $\delta n_{l,m} = a_{l,m}(x_0)$ for $l \ge 1$. The two equations for $l=1$ and $l=0$ read 
\bea
	x_0 \delta n_{1,0} &=& \frac{1+F_0}{\sqrt 3} \delta n_{0,0} + \frac2{\sqrt{15}} \delta n_{2,0},  \\
	x_0 \delta n_{0,0} &=& \frac{1}{\sqrt{3}} \delta n_{1,0},
\eea
which lead to the following solution,
\bea
	\delta n_{0,0} &=&  2\sqrt \pi (x_0 \arccoth x_0 -1), \\
	\quad x_0 \arccoth x_0  &=& 1+  \frac1{F_0}.
\eea
This is of course consistent with the previous eigen-solution~(\ref{eq:n0}).

To access the hydrodynamic regime, where the collisionless zero sound crosses over to the hydrodynamic first sound, the collision integral plays an essential role. We take the collision integral to have the following form~\cite{Lucas:2018electronic}
\bea
	\I(n) = -   \sum_{l=0}^\infty \gamma_l \delta n_{l,0}, \quad 	\gamma_l = \begin{cases}
				0 & l = 0\\
				\gamma_\text{imp} & l = 1 \\
				\gamma & l \ge 2
				\end{cases}.
\eea
where $\gamma$ is the collision rate from collisions between quasiparticles and $\gamma_\text{imp}$ is the collision rate between quasiparticles and impurities. Obviously, $\gamma$ is the key hydrodynamic interaction parameter. Because collisions between quasiparticles conserve the particle number and their total momentum, the pure quasiparticle collision rate for $l=0, 1$ vanishes by virtue of conservation laws. On the other hand, elastic collisions between quasiparticles and quenched impurities relax the momentum, and, therefore, $\gamma_\text{imp}$ is nonzero for $l=1$. With this collision integral, the Boltzmann equation in the basis of spherical harmonics reduces to the following coupled equations,
\bea
	x \delta n_{l,0} &=& \frac{l}{\sqrt{4l^2-1}} \delta n_{l-1,0} + \frac{l+1}{\sqrt{4(l+1)^2-1}} \delta n_{l+1,0} , \nn\\
	 && \qquad\qquad\qquad\qquad\qquad\qquad\qquad l \ge 2, \\
	x_\text{imp} \delta n_{1,0} &=& \frac{1 + F_0}{\sqrt 3} \delta n_{0,0} + \frac2{\sqrt{15}} \delta n_{2,0},  \\
	x_0 \delta n_{0,0} &=& \frac{1}{\sqrt{3}} \delta n_{1,0}, \\
\label{eq:x}	x_0 &=& \frac{\omega}{q v_F},~ x = \frac{\omega + \ii \gamma}{q v_F}, ~ x_\text{imp} = \frac{\omega + \ii \gamma_\text{imp}}{q v_F}.
\eea
The series~(\ref{eq:series}) again solves the first recurrence relation, with $x$ replacing $x_0$. Thus for $l \ge 1$, $\delta n_{l,m} = a_{l,m}(x) $. The other two equations can be solved easily, namely, $\delta n_{0,0} = 2 \sqrt \pi \frac{x}{x_0} (x \arccoth x - 1)$ and
\bea \label{eq:eigen}
 (F_0 + 1 -3 x_0 x_\text{imp}) (x \arccoth x - 1) \nn\\ + x_0 [(3x^2-1) \arccoth x - 3x ] = 0.
\eea
This is the main result of our paper. The eigen dispersion of the sound mode is determined by~(\ref{eq:eigen}), where the definitions of $x$'s are given by~(\ref{eq:x}).

To investigate the sound mode, we can assume $\gamma_\text{imp} = 0$, otherwise the coherent propagating sound mode is damped. In the collisionless regime, $\omega \gg \gamma$, we obtain the zero sound mode or the plasmon in~(\ref{eq:zeroFull}) as
\bea
	&& \omega = \pm v_0 q - \ii \gamma \frac{(F_0+1)^2 - 2(F_0-1) (\frac{v_0}{v_F})^2 - 3 (\frac{v_0}{v_F})^4 }{F_0 [ F_0 +1 - (\frac{v_0}{v_F})^2]}, \\ && \frac{v_0}{v_F} \arccoth \frac{v_0}{v_F} = 1 + \frac1{F_0}.
\eea
In the case of strong repulsion $F_0 > 1$ (this is also the case for Coulomb interaction in the long wavelength limit), the sound mode can be simplified to
\bea
	\omega = \pm \sqrt{\frac{F_0}3 + \frac35} v_F q - \ii \gamma \frac{2(5F_0 + 21)}{5F_0(5F_0 + 3)},
\eea
which is nothing but~(\ref{eq:zero}). On the other hand, in the collision-dominated hydrodynamic regime $\omega \ll \gamma$ we get the first sound,
\bea
	\omega = \pm \sqrt{ \frac{F_0}3 + \frac13} v_F q - \ii \frac{2 v_F^2}{15 \gamma} q^2,
\eea
which is~(\ref{eq:first}).

We can also consider the effect of the finite impurity scattering. For weak impurity scattering, the effect is to modify the damping defined by imaginary parts of the sound mode. For the zero sound, we have
\bea \label{eq:weak_imp}
\omega &=& \pm v_0 q - \ii \gamma \frac{(F_0+1)^2 - 2(F_0-1) (\frac{v_0}{v_F})^2 - 3 (\frac{v_0}{v_F})^4 }{F_0 [ F_0 +1 - (\frac{v_0}{v_F})^2]} \nn\\
&& - \ii \gamma_\text{imp} \frac{3(\frac{v_0}{v_F})^2[(\frac{v_0}{v_F})^2-1]}{F_0(F_0 +1 -(\frac{v_0}{v_F})^2)}.
\eea
For large $F_0 > 1$, this reduces to a simple correction $\delta \Im(\omega) = \frac12 \gamma_\text{imp}$. 
For strong impurity scattering $\gamma_\text{imp} > q$, both sound modes are over-damped into
\bea
	\omega = - \ii \frac{F_0+1}{3\gamma_\text{imp}} v_F^2 q^2, \qquad \omega = -\ii \gamma_\text{imp}.
\eea

\section{Conclusions} \label{sec:conclusion}

We have discussed electronic sound modes in 3D metals in the presence of long-range Coulomb coupling via the Boltzmann equation. A more microscopic approach to the collective mode, like plasmons, would be to start from the electron Hamiltonian with long-range interactions. Then the collective mode results from integrating out the electron fluctuations. In the lowest order, this is nothing but the RPA approach. Here, we recapitulate how it works for 3D metals. In the RPA approximation, the collective mode is determined by the following eigen equation,
\bea \label{eq:RPA}
	1- V(\tb q) \Pi(\tb q, \omega) = 0,
\eea
where $V(\tb q)$ is the interaction at momentum $\bf q$, and the dynamical electron polarization function is defined by
\bea
	&& \Pi(\tb q, \omega) = \frac{1}{\beta} \sum_n \int_{\tb k} \frac1{-\ii \Omega_n - \frac\omega2+ \xi(\tb k + \frac{\tb q}2)} \frac1{-\ii \Omega_n + \frac\omega2+ \xi(\tb k - \frac{\tb q}2)}, \nn \\
	\\
	&& \Omega_n = \frac{(2n+1)\pi}{\beta}, \quad \xi(\tb k) = \epsilon_0(\tb k) - \mu.
\eea
where $\Omega_n$ is the Matsubara frequency and $\omega$ is the real frequency. To proceed, we assume a spherical Fermi surface with parabolic dispersion $\epsilon_0(\tb k) = \frac{\tb k^2}{2m}$. It is not hard to get the 3D polarization function at zero temperature, namely,
\bea \label{eq:polarization3D}
	\Pi(\tb q, \omega) = \frac{N_e q^2}{m \omega^2} \left( 1 + \frac35 \frac{k_F^2}{m^2} \frac{q^2}{\omega^2} \right), \quad N_e = \frac{4\pi}3 \frac{k_F^3}{(2\pi)^3},
\eea
where $q \equiv |\tb q|$. Putting in the Coulomb potential given by $V_\text{Cou}(q) = \frac{e^2}{q^2}$ and the polarization function~(\ref{eq:polarization3D}) into eigen equation~(\ref{eq:RPA}), we obtain the conventional plasmon given in~(\ref{eq:plasmon3D}). 

We can also consider a short-range density-density interaction potential that is independent of momentum, namely,
\bea 
	V({\bf q}) = \frac{2\pi^2 v_F}{ k_F^2} F_0,
\eea
where the prefactor originates from how the Landau parameter is introduced via quasiparticle interactions at the Fermi surface, i.e., $\int_{F.S.} \frac{d^3 {\bf q}}{(2\pi)^3}  V({\bf q})...= \int \frac{d\Omega}{4\pi} (\frac{ k_F^2}{2\pi^2 v_F} V(\Omega))... $. For the s-wave channel, it reduces to the above equation.
Using this interaction potential to replace the Coulomb potential, we get the following linearly dispersing sound-like collisionless collective mode 
\bea
	\omega^2 = \frac{v_F F_0}{4\pi k_F^2} \frac{N_e q^2}{m} \left( 1 + \frac35 \frac{k_F^2}{m^2} \frac{q^2}{\omega^2} \right), \quad \omega \approx \pm \sqrt{\frac{F_0}3 + \frac35} v_F q. \nn \\
\eea
The second equation is exactly the zero sound mode in~(\ref{eq:zero}). At zero temperature, within the RPA, there is no damping of the electronic collective mode by quasiparticle collisions since the collision rate vanishes as $T^2$, but finite impurity scattering would still contribute to the damping in the way we discussed earlier.  At finite temperatures, quasiparticle collisions would lead to Landau damping of the collective modes. 

This simple calculation tells us that microscopically the interaction potential determines the dispersion of the electronic collective modes. The zero sound mode for short-range interactions becomes the gapped 3D plasmon mode in the presence of long-range Coulomb coupling. (The first sound mode also acquires a long wavelength gap as discussed earlier.) Indeed, from the perspective of symmetry, both sound modes and plasmons are density fluctuations that characterize the underlying particle number conservation. So they are actually the same collective mode, and it is not a surprise that they all develop long wavelength gaps because of the long-range Coulomb interaction. Thus, the hydrodynamic sound in 3D metals is not a sound mode at all since it has a finite energy at zero momentum defined by the 3D plasma frequency. We do note, however, that the sound modes differ from the plasmon in their next-to-leading-order dispersion corrections at finite momentum. 
Although the 2D case is clearly addressed in Ref.~[3], 
we briefly discuss the RPA calculation in 2D for completeness. 
The electron polarization function in two dimensions reads
\bea \label{eq:polarization2D}
	\Pi({\bf q}, \omega) = \frac{k_F^2 q^2}{4 \pi m \omega^2} \left( 1 + \frac34 \frac{k_F^2}{m^2}\frac{q^2}{\omega^2} \right).
\eea
Putting the short-range potential given by $V({\bf q}) = \frac{2\pi v_F}{ k_F} F_0$ and the polarization function~(\ref{eq:polarization2D}) into the eigen equation~(\ref{eq:RPA}), we recover the linear zero sound $\omega = \pm (\frac{F_0}{2} + \frac34 )^{1/2}v_F q$.
On the other hand, using instead the 2D Coulomb potential given by $V_\text{Cou}({\bf q}) = \frac12 \frac{e^2}{|{\bf q}|}$ (the factor of $\frac12$ comes from the two dimensional Fourier transform of the usual Coulomb potential $V_\text{Cou}({\bf r}) = \frac1{4\pi} \frac1{|\bf r|}$) we obtain the following collective plasmon mode
\bea	
	\omega = \pm \sqrt{\frac{\pi \alpha v_F^2}{\lambda_F} |\bf q|}.
\eea
Therefore, within the RPA calculation, one can already see that the linear sound wave is modified to be $\omega \propto \sqrt{q}$ by replacing the short-range interaction by the long-range Coulomb interaction. Going beyond the RPA framework, it was shown in Ref.~[3] 
that although both the first and second sound waves exhibit the same plasmon-like dispersion, the next-to-the-leading order momentum dependence differs.

With this in mind, we now briefly discuss the 1D case.  In 1D, the Boltzmann approach does not work simply because the Fermi liquid does not exist in the presence of any finite interaction~\cite{Tomonaga:1950remarks, Luttinger:1963exactly}. 
Thus, starting from the electron Hamiltonian including a single-particle dispersion (we take a parabolic dispersion for simplicity) and interaction potentials $V(\bf q)$ like that given above is a good and simple way to look for sound or plasmon modes. 
The electron polarization function is now given by~\cite{Sarma:1985screening}
\bea
	\Pi(q, \omega) = \frac{m}{2\pi q} \ln \left( \frac{m^2 \omega^2 - (k_F- \frac{q}2)^2 q^2}{m^2 \omega^2 - (k_F+ \frac{q}2)^2 q^2} \right).
\eea
We expect a sound wave-like (linear in momentum) mode when the interaction is short ranged. Indeed the short-range potential $V(q) = \pi v_F F_0$ leads to the zero sound mode in 1D,
\bea
 	\omega = \pm v_F \sqrt{F_0} q.
\eea
How does the Coulomb interaction affect this sound mode? The answer is simple, we just need to replace $V(q)$ with the 1D Coulomb potential~\cite{Sarma:1985screening},
\bea
	V_\text{Cou}(q) = \frac{e^2}{4\pi} \int dr \frac{e^{\ii q r}}{\sqrt{r^2 + a^2}} = \frac{e^2}{4\pi} 2 K_0(a q),
\eea
where $K_0(x)$ is the modified Bessel equation of the second kind and $a$ is a short-range cutoff introduced to make the integral converge in 1D ($a$ can be thought of as a lattice constant). Plugging the Coulomb potential and the polarization function in 1D into the eigen equation~(\ref{eq:RPA}), the resultant long wavelength collective mode is 
\bea
	\omega = \pm \frac{e}{\pi} \sqrt{\frac{v_F}2} q \sqrt{- \ln \frac{a q}2}, \quad a q \ll 1,
\eea
which is nothing but the well-known plasmon mode in 1D. Although we consider zero temperature, we expect that the plasmon mode takes over the sound modes in 1D when the Coulomb interaction dominates since the dispersion is largely independent of temperature. A more physical argument is that from the symmetry perspective, sound modes and plasmons are the very same mode, and the different names just refer to whether the interaction potential is short-range or long-range. The curious thing is that in 3D systems, where the Coulomb coupling goes as $q^{-2}$, the sound mode is not acoustic at all since it acquires the plasmon mass at zero momentum.

In conclusion, we constructed a solvable model in 3D to obtain the sound modes in both the collisionless regime and the collision-dominated hydrodynamic regime. In particular, we discussed the effect of long-range Coulomb interaction on the sound modes. We found that in the presence of Coulomb interaction, both the zero sound and the first sound obtain a finite gap equal to the plasmon frequency, and a damping rate which is quadratic in momentum. We also discussed general long-range interactions that lead to unusual plasmon dispersions. Finally, we clarified the collective mode and sound mode dichotomy in 1D.

\section*{Acknowledgments}

This work is supported by the Laboratory for Physical Sciences. S.-K.J. is supported by the Simons Foundation via the It From Qubit Collaboration.

\appendix

\begin{widetext}

\section{Coulomb potential} \label{append:Coulomb}

In this appendix, we present the Fourier transform of the 3D Coulomb potential. Since the Coulomb potential is spherically symmetric, without loss of generality, we can choose the momentum pointing to the $z$ axis and make a coordinate transform as follows,
\bea
	\int d^3 \tb r \frac{1}{|\tb r |} e^{\ii \tb q \cdot \tb r} &=& \int dr d\theta d\phi r^2 \sin \theta \frac1r e^{\ii |\tb q| r \cos \theta - \varepsilon r} \\
	&=& \int dr \frac{4\pi e^{-\varepsilon r }\sin (|\tb q| r)}{|\tb q|} = \frac{4\pi}{\tb q^2 + \varepsilon^2}.
\eea 
where in the second step we have added an infinitesimal positive number $\varepsilon$ to ensure the convergence. In the last step, we can safely set $\varepsilon$ to zero.

\section{Some useful mathematical results} \label{append:math}

In this appendix we present mathematical results that are used in the main text. The spherical harmonics are defined by
\bea
	Y_l^m(\theta, \phi) = \sqrt{\frac{(2l+1)(l-m)!}{4\pi (l+m)!}} P_l^m(\cos \theta) e^{\ii m \phi},	\quad |m| \le l, \quad l \ge 0, 
\eea
where $P_l^m(x)$ is the associated Legendre polynomial, and the prefactor is chosen to make sure the spherical harmonics are properly normalized,
\bea
	\int d\Omega Y_l^m(\theta, \phi) Y_{l'}^{m'}(\theta, \phi)^\ast = \delta_{ll'} \delta_{mm'},
\eea
where $d \Omega \equiv \sin\theta d \theta d\phi$ is the short-hand notation for the measure of a sphere. A useful recurrence formula of the associated Legendre polynomial is
\bea
	 x P_l^m(x) = \frac{l-m+1}{2l+1} P_{l+1}^m(x) + \frac{l+m}{2l+1} P_{l-1}^m(x).
\eea
With this recurrence relation, we arrive at the following recurrence formula 
\bea
	\cos \theta Y_l^m(\theta, \phi) = \sqrt{\frac{(l+1)^2-m^2}{4(l+1)^2-1}} Y_{l+1}^m(\theta, \phi) + \sqrt{\frac{l^2-m^2}{4l^2-1}} Y_{l-1}^m(\theta, \phi).
\eea
This is useful to transform the Boltzmann equation into the harmonic basis.

We are interested in the expansion of the following function in the basis of spherical harmonics,
\bea
	\delta n_0(\Omega) = \frac{\cos \theta}{\lambda - \cos \theta}.
\eea
Since this function is independent of the angle $\phi$, its expansion coefficient on $Y_l^{m \ne 0}(\theta, \phi)$ vanishes. We can focus on the $Y_l^0(\theta, \phi)$ which is related to the Legendre polynomial $P_l(x)$ (By definition, the Legendre polynomial is a specific case of the associated Legendre polynomial $P_l(x) \equiv P_l^{m=0}(x)$). Let us first evaluate the integral
\bea
	\int_{-1}^1 dx \frac{x}{\lambda - x} P_n(x) = \frac1{2^n n!} \int_{-1}^1 dx \frac{x}{\lambda - x} \frac{d^n}{dx^n} (x^2 -1)^n, \nn \\
\eea
where in the first step we have performed a coordinate transformation $\cos \theta = x$, and in the second step we have used Rodrigues's formula  $P_n(x) \equiv \frac1{2^n n!}  \frac{d^n}{dx^n} (x^2 -1)^n $. For $n=0$, we can directly evaluate the integral to get
\bea
	\int_{-1}^1 dx \frac{x}{\lambda - x} P_0(x) = 2 (\lambda \arccoth \lambda -1).
\eea

For $n \ge 1$, we can repeatedly use integration by parts to bring the integral into the form
\bea
	\int_{-1}^1 dx \frac{x}{\lambda - x} P_n(x) &=& \frac1{2^n n!} \int dx \left( \frac{d^n}{dx^n} \frac{x}{\lambda-x}\right) (1-x^2)^n + [\text{boundary terms}].
\eea
It is not hard to show that all boundary terms vanish. The $n$-th derivative of the function $\delta n_0(\Omega)$ is 
\bea
	\frac{d^n}{dx^n}	 \frac{x}{\lambda - x} = \frac{n! \lambda}{(\lambda - x)^{n+1}}.
\eea
Plugging this into the integral, we have
\bea
	&& \int_{-1}^1 dx \frac{x}{\lambda - x} P_n(x) \\
	&=& \frac{\lambda}{2^n} \int_{-1}^1 dx \frac{(1-x^2)^n}{(\lambda - x)^{n+1}} \\
	&=& \frac{\sqrt \pi}{(2\lambda)^n} \frac{\Gamma(n+1)}{\Gamma(n+ \frac32)} \ _2F_1\left( \frac{n+1}2 , \frac{n+2}2; n + \frac32; \frac1{\lambda^2}\right),
\eea
where $\ _2F_1(a,b;c;z)$ denotes the hypergeometric function. Finally, in terms of the spherical harmonics,
\bea
	&& \int d\Omega \frac{\cos \theta}{\lambda -\cos \theta} Y_l^m(\theta, \phi)^\ast \\
	&=&  2\pi \delta_{m,0} \sqrt{\frac{2l+1}{4\pi}} \int_0^\pi d\theta \sin \theta \frac{\cos \theta}{\lambda -\cos \theta} P_l(\cos \theta) \\
	&=& 2\pi \delta_{m,0} \sqrt{\frac{2l+1}{4}}  \frac{1}{(2\lambda)^l} \frac{\Gamma(l+1)}{\Gamma(l+ \frac32)} \ _2F_1\left( \frac{l+1}2 , \frac{l+2}2; l + \frac32; \frac1{\lambda^2}\right).
\eea
The expansion coefficient of $\delta n_0(\Omega) = \sum_{l=0}^\infty \sum_{m=-l}^l \delta n_0^{lm}(\lambda) Y_l^m(\theta, \phi) $ is
\bea
	\delta n_0^{lm}(\lambda) = \begin{cases} 
	\delta_{m,0} \times 2\sqrt \pi (\lambda \arccoth \lambda - 1)  & l=0	\\
	 \delta_{m,0} \times 2\pi  \sqrt{\frac{2l+1}{4}}  \frac{1}{(2\lambda)^l} \frac{\Gamma(l+1)}{\Gamma(l+ \frac32)} \ _2F_1\left( \frac{l+1}2 , \frac{l+2}2; l + \frac32; \frac1{\lambda^2}\right) & l \ge 1
	 \end{cases}.
\eea

\end{widetext}

\bibliography{references}

\end{document}